\documentclass[useAMS,usenatbib,psfig]{mn2e}
\usepackage{graphicx}
\begin{document}
\title[The 2010 outburst of V407 Cyg]{The 2010 nova outburst of the symbiotic Mira V407 Cyg\thanks{Tables 1-4 available in electronic form only}}
\author[U. Munari et al.]{U. Munari$^{1}$, V.H. Joshi$^{2}$, N.M. Ashok$^{2}$, D.P.K. Banerjee$^{2}$, P. Valisa$^{3}$, A. Milani$^{3}$,
\newauthor A. Siviero$^{1}$, S. Dallaporta$^{3}$, F. Castellani$^{3}$\\
$^{1}$INAF Astronomical Observatory of Padova, 36012 Asiago (VI), Italy\\
$^{2}$Astronomy and Astrophysics Division, Physical Research Laboratory, Navrangapura, Ahmedabad - 380009, Gujarat, India\\
$^{3}$ANS Collaboration, c/o Osservatorio Astronomico, via dell'Osservatorio 8, 36012 Asiago (VI), Italy}

\date{Accepted .... Received ....; in original form ....}

\pagerange{\pageref{firstpage}--\pageref{lastpage}} \pubyear{2010}

\maketitle

\label{firstpage}

\begin{abstract}
The nova outburst experienced in 2010 by the symbiotic
binary Mira V407 Cyg has been extensively studied at optical and infrared
wavelengths with both photometric and spectroscopic observations.  This
outburst, reminiscent of  similar events displayed by RS Oph,
can be  described as a very fast He/N nova erupting while being deeply
embedded in the dense wind of its cool giant companion.  The hard
radiation from the initial thermonuclear flash  ionizes and excites
the wind of the Mira over great distances (recombination is observed on a time
scale of 4 days). The novae ejecta is found to progressively decelerate
with time as it expands into the Mira wind. This is deduced from line widths which 
change from  a FWHM of 2760 km~s$^{-1}$ on day +2.3 to 200 km~s$^{-1}$
on day +196.  The wind of the Mira is massive and extended enough for an
outer neutral and unperturbed region to survive at all outburst phases.
\end{abstract}

\section{Introduction}

The symbiotic binary V407 Cyg consists of an accreting white dwarf and an
O-rich Mira companion pulsating with a 745 day period.  Miras with such a
long pulsation period are generally OH/IR sources with a very thick dust
envelope which prevents direct observation of the central star at optical
wavelengths.  The much thinner dust envelope in V407 Cyg is probably due to
the presence of the WD companion whose orbital motion, hard radiation field
in quiescence and violent mass ejection during outbursts inhibits dust
formation in a large fraction of the Mira wind (Munari et al.  1990,
hereafter M90).

V407 Cyg was discovered by Hoffmeister (1949) as Nova Cyg 1936, just at the
time when its Mira was passing through maximum brightness.  No spectroscopic
observations confirming it as a genuine nova outburst were however
available.  What actually occured is unclear because (i) the object was
discovered and remained at B$\approx$14.5 mag for an entire Mira pulsation
cycle, without declining to an expected $B$$\geq$19 minimum (cf Figure 1 in
M90), but at the same time (ii) the peak brightness was much smaller than
$B$$\sim$8 reached by V047 Cyg during its present 2010 outburst which
resembles a true nova eruption.  The 1936-1938 event could have been one of
the usual low amplitude, long-lasting outbursts that symbiotic binaries
frequently display.  Two such active phases during the 1990's were reported
and discussed by Munari et al.  (1994), Kolotilov et al.  (1998, 2003,
hereafter K98 and K03), and some earlier ones can be spotted in the
historical light-curves of V407 Cyg by Munari and Jurdana-\v{S}epi\'{c}
(2002) and M90.

The 2010 outburst of V407 Cyg was discovered on March 10.813 UT by Nishiyama
and Kabashima (2010) at $V$=7.6 mag.  This was at an unsurpassed brightness
level in the star's recorded photometric history thereby underscoring the
peculiarity and importance of the event.  The first spectroscopic
confirmation and analysis of the outburst was given by Munari et al. 
(2010a) who described the event as a He/N nova expanding within the wind of
the Mira companion.  The similarity with RS Oph was also pointed out. 
In the following weeks and months the outburst was intensively monitored
over several wavelength regimes viz.  in $\gamma$-rays (Abdo et al.  2010,
Cheung et al.  2010), radio (Krauss et al.  2010, Giroletti et al.  2010,
Bower et al.  2010, Nestoras et al.  2010, Gawronski et al.  2010, Pooley
2010), SiO maser (Deguchi et al.  2010), and infrared (Joshi et al.  2010).

So far, apart from brief circulars, no comprehensive report on the
photometric and spectroscopic evolution of V407 Cyg at optical and IR
wavelengths is available.  The aim of this Letter is thus to provide a first
report; follow-up papers will present a more detailed analysis and modeling
of the huge amount of data we have and are still collecting.

  \begin{figure*}
     \centering
     \includegraphics[height=17.0cm,angle=270]{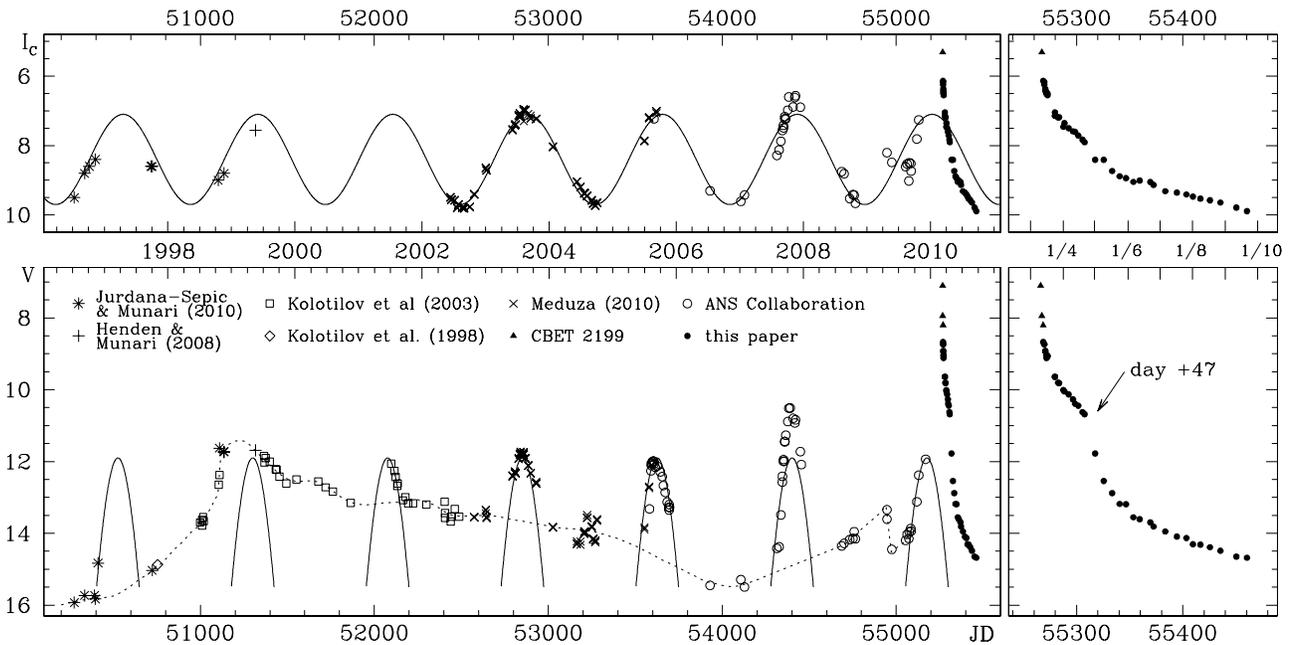}
     \caption{Photometric evolution of V407 Cyg over the last
     16 years, or 7 Mira's pulsation cycles. The dashed line is hand-drawn
     to provide a guide through the long-lasting active phase that peaked
     in 1998. The right panels provide a zoom over the 2010 outburst.}
     \label{fig1}
  \end{figure*}

\section{Observations}

Optical photometry was recorded with several small telescopes operated by the
ANS Collaboration in northern Italy, all equipped with CCDs and photometric
{\em UBVR$_{\rm C}$I$_{\rm C}$} filters.  Corrections for bias, dark, and
flat fields were applied in the usual manner.  Photometric calibration and
correction for color equations was performed for all instruments against the
same {\em UBVR$_{\rm C}$I$_{\rm C}$} sequence calibrated by Henden and
Munari (2000) around V407 Cyg.  Our photometry of V407 Cyg covering the 2010
outburst is presented in Table~1 and plotted in Figure~1 and 2.

Optical spectroscopy was obtained with different telescopes:
Asiago 1.82m + Echelle spectrograph (20000 resolving power), Asiago
1.22m + B\&C spectrograph (low resolution mode), Varese 0.6m + multi
mode spectrograph. A journal of the observations is given in Table~2.
With the Varese 0.6m telescope we obtained both low resolution and Echelle
spectra. The latter were recorded both in unbinned (resolving power
17000, marked {\em ech} in Table 2) and binned mode (resolving power
10000, marked {\em echB} in Table 2). All spectra (including Echelle
ones) were calibrated in absolute fluxes by observations of several
spectrophotometric standards during the night. Their zero-points were
then checked against simultaneous BVRI photometry by integrating the
band transmission profiles on the fluxed spectra.

Near-IR observations were carried out in the $J, H, K$ bands at the Mt.
Abu 1.2m telescope during the early outburst phase. The spectra were
obtained at a resolution of $\sim$$1000$ using a NICMOS3
Imager/Spectrometer. Spectra and photometry of the comparison star HR 7984
were also obtained for the spectro-photometric data reduction.
Wavelength calibration was done using OH sky lines and telluric features
that register with the stellar spectra. The detailed reduction of the
spectral and photometric data, using IRAF tasks, follow a standard
procedure that is described for e.g in Naik et al. (2009). The journal of
infrared spectroscopic observations is given in Table~3, and the results of
infrared photometry in Table~4 and Figure~2.

 \begin{figure}
 \centering
  \includegraphics[width=8.3cm]{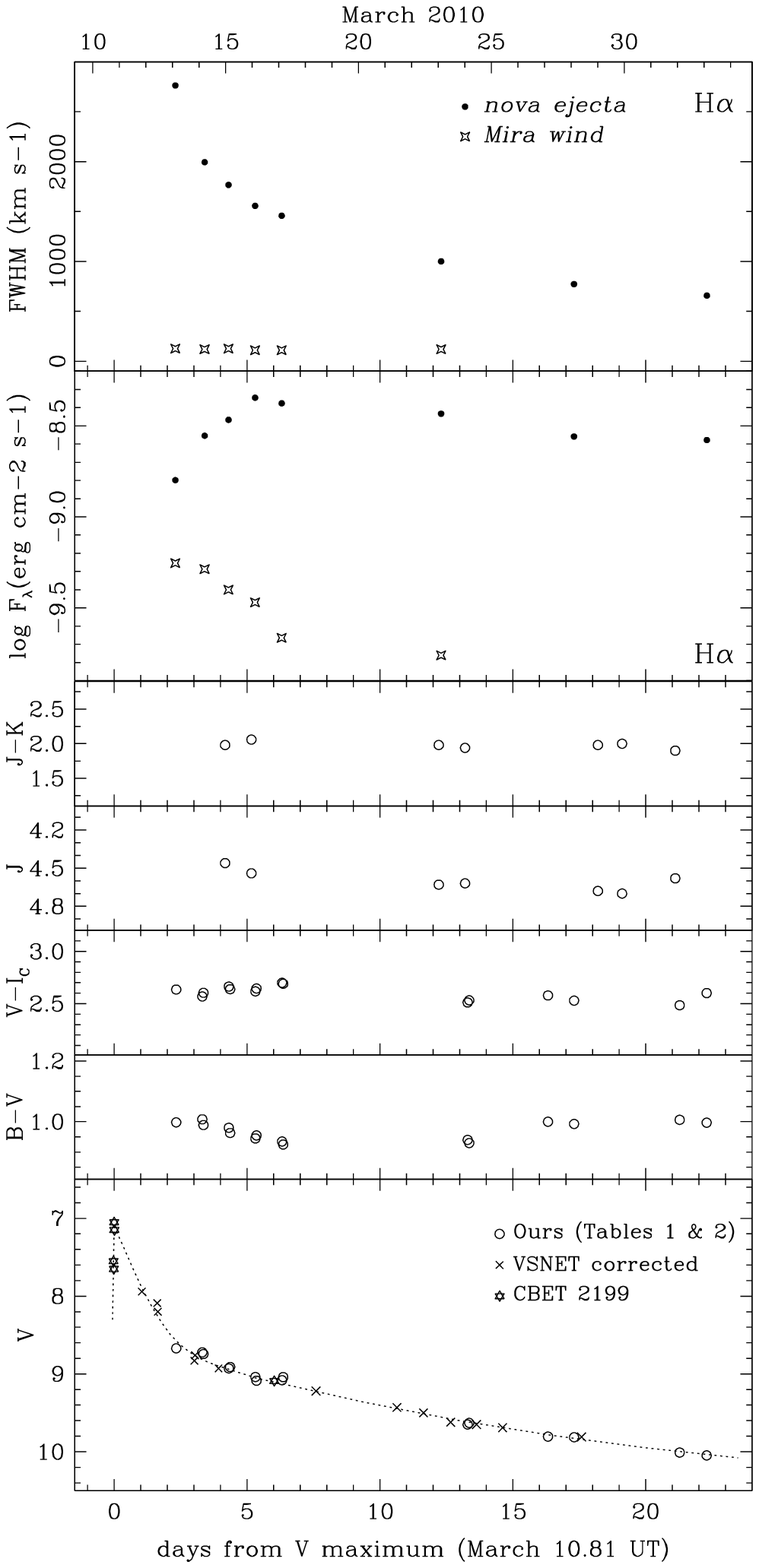}
  \caption{{\em Lower panels}: early optical and infrared photometric
     evolution of V407 Cyg during the 2010 outburst.  The dashed line is
     hand-drawn for guidance.  {\em Upper panels}: evolution in
     width and integrated flux of the H$\alpha$ emission line.  `Nova
     ejecta' refer to the broad component (cf. sect.  3), `Mira wind'
     to the superimposed narrow one (see Figure 5).}
     \label{fig2}
  \end{figure}

\section{Results}

The observations presented in this paper show that the violent outburst
experienced by V407 Cyg in March 2010 was a thermonuclear runaway (TNR), the
same event that powers a normal nova eruption.  In normal novae, the ejected
material essentially expands freely into a void circumstellar medium. 
However, in V407 Cyg, the fast ejecta have to expand into the dense and slow
wind of the Mira companion, and are thus progressively slowed down as the
pre-existing circumstellar material is swept up in an expanding shell. 
Noteworthy, the pre-existing circumstellar material offers an ideal
ionization target for the hard radiation from the initial TNR flash.

The similarity with the outburst displayed by the celebrated RS Oph is
evident (Bode 1987, Evans et al.  2008, and references therein).  The
latter is a symbiotic binary, with an orbital period of 460 days and an M
giant filling its Roche lobe (Schaefer 2009) which transfers material to a
massive WD (Hachisu et al.  2007).  Similar nova eruptions have been seen
also in the symbiotic binaries and recurrent novae T CrB, V745 Oph and V3890
Sgr (Schaefer 2010).

What occured in V407 Cyg is well illustrated by the evolution of the
H$\alpha$ profile (Figure~5) and its width and integrated
flux (top panels of Figure~2). At the earliest
stages, the H$\alpha$ profile is dominated by a sharp component
superposed on a much broader one, as first noted by Munari et al. (2010a).

The sharp component, identical to that in quiescence but enormously brighter
(cf profiles for 2008 and 2009 in Figure 5), is due to the sudden ionization
of a large fraction of the Mira's wind by the flash of energetic radiation
produced by the TNR event.  The wind of the Mira does not as yet get perturbed
kinematically, as proven by the preserved sharpness of the H$\alpha$ profile
that increased its emissivity by two orders of magnitude compared to
quiescence.  The flux of hard photons, however, is not large enough to
ionize the whole Mira wind, as indicated by the persistence of the sharp
absorption component which maintains the same heliocentric radial velocity
as in quiescence ($-$50 km/sec). The intensity of the H$\alpha$ sharp
component rapidly declines subsequently (cf Figure~2), with a recombination time scale
of 4 days, which can be written as
\begin{equation}
t_{\rm rec} = 0.66 \left(\frac{T_{\rm e}}{10^4 {\rm ~K}}\right)^{0.8}
\left(\frac{n_{\rm e}}{10^9 {\rm ~cm}^{-3}}\right)^{-1} \approx 100 {\rm ~hours}
\end{equation}
following Ferland (2003). It corresponds to a density of about 
5$\times$10$^{6}$ cm$^{-3}$ for the fraction of the Mira wind ionized by the TNR initial
flash.  The point at day +12.3 in Figure~2, e.g.  the last epoch at which a narrow
component could still be resolved in the H$\alpha$ profiles of Figure~5,
deviates from the $t_{\rm rec}$=4 days of earlier points.  By this time, the
nova ejecta has begun to turn  optically thin and the hard radiation field of the
central star (presumably still burning hydrogen at its surface during the
constant luminosity phase) is hot and intense enough to produce coronal
emission lines, as reported by Munari et al. (2010b). The same radiation field,
leaking through the optically thin ejecta, is also responsible for
sustaining the ionization of the circumstellar gas not yet reached by the
expanding shell.

  \begin{figure*}
     \centering
     \includegraphics[width=7.0cm,height=17cm,angle=270]{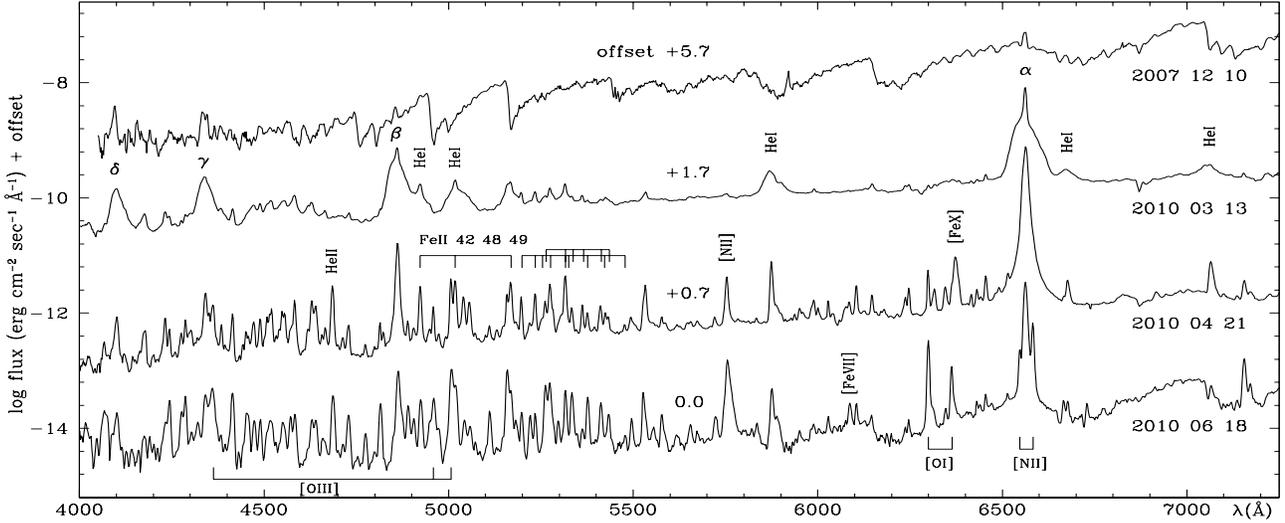}
             \caption{A sample of the absolutely fluxed, optical spectra of V407 Cyg
     that we collected during the 2010 outburst. The 10 December 2007
     spectrum shows the quiescence spectrum at the time of Mira brightness
     maximum. Only some of the emission lines are identified.}
     \label{fig3}
  \end{figure*}

  \begin{figure*}
     \centering
     \includegraphics[width=7.0cm,height=17cm,angle=270]{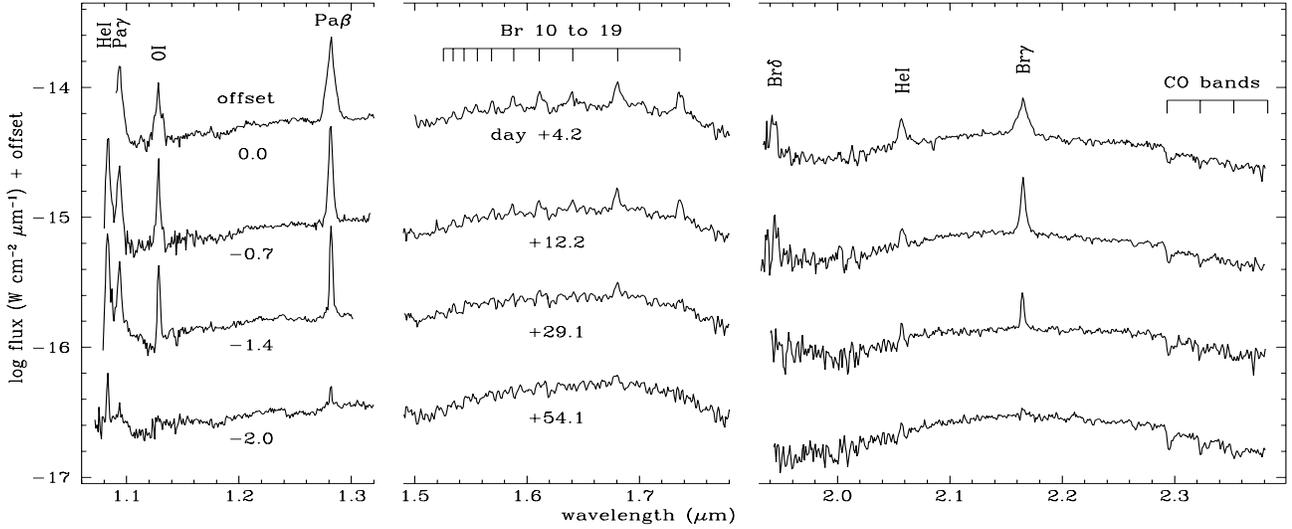}
     \caption{A sample of our absolutely fluxed, infrared spectra of V407 Cyg
     for the 2010 outburst. Major emission lines are identified.}
     \label{fig3}
  \end{figure*}

The broad component of the V407 Cyg H$\alpha$ profiles in Figure~5,
originates instead in the material ejected at high velocity, as in any
normal nova.  The broad spectrum nicely matches that of a normal "He/N" nova
(Williams 1992) as  illustrated by the low resolution optical and
infrared spectra for days +2.3 and +4.2 in Figure~3 and 4 respectively.  A
He/N spectrum is typical of fast novae and of RS Oph too.  The nova ejecta
is rapidly decelerated while trying to expand through the surrounding Mira
wind and the distinction between a sharp and a broad component to the
emission lines is then progressively attenuated, disappearing two weeks
past optical maximum.  As more material is swept by the expanding shell,
the velocity continues to decrease.  Figure~2 illustrates the temporal evolution 
of the FWHM (in km sec$^{-1}$) of the broad component of H$\alpha$,
which is accurately fitted by the expression
\begin{equation}
{\rm FWHM} = 4320 - 5440 \log t + 2635 (\log t)^2 - 460 (\log t)^3
\end{equation}
including later phases characterized by 400, 280 and 200 km~s$^{-1}$ on days
 +48.2, +105 and +196 respectively. The same trend is shared also by the hydrogen
lines dominating the infrared spectra of Figure~4. For comparison the FWHM of H$\alpha$
in quiescence was stable at $\sim$120 km~s$^{-1}$ (cf profiles for 2008 and 2009 in Figure 5).
Figure~5 shows the emergence of [NII] 6548, 6584 \AA\ doublet two months
past optical maximum.  It did not originate in the expanding material, but
instead in the outer wind of the Mira, external to the expanding shell.
This is proved since its profile FWHM of $\sim$110 km~s$^{-1}$  is
much sharper than that of the adjacent H$\alpha$ and identical to the width in
quiescence. The existence of an outer region of the Mira wind not yet reached
on day +196 by the already greatly slowed down ejecta (cf the sharp absorption
component at $-$50 km s$^{-1}$ in Figure~5), leads us to speculate that
some part of the ejecta could remain bound to the binary system and
could be re-accreted at later times by the WD.

  \begin{figure*}
     \centering
     \includegraphics[height=17.0cm,angle=270]{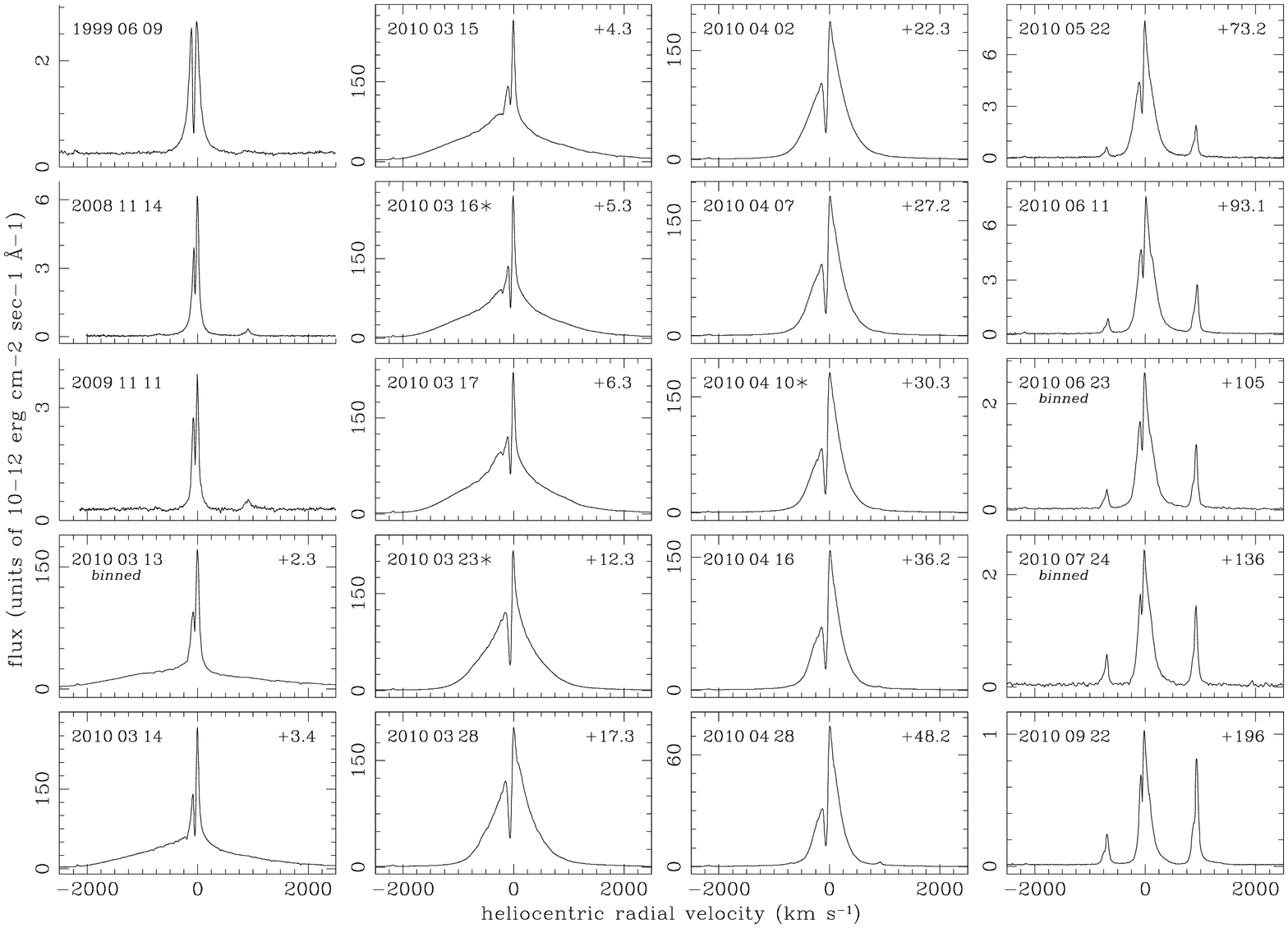}
     \caption{Evolution of the H$\alpha$ profile of V407 Cyg during the 2010
     outburst.  Older profiles are given for reference purposes, and pertain
     to the brightness peak during the 1997-2006 active phase (1999),
     and to quiescence at the time of a Mira minimum (2008) and maximum
     (2009). The three spectra marked with $\ast$ are courtesy of Christian Buil
     and fluxed by us against our low-resolution spectra.}
     \label{fig3}
  \end{figure*}

The light-curve of V407 Cyg over the last 15 years is presented in Figure~1.
It is characterized by three main components: (1) the 745 day pulsation of
the Mira (sinusoid drawn as a solid line), which dominates the light-curve
at reddest wavelengths; (2) the presence of a limited amplitude, slow
evolution active phase (dashed line in the $V$ band panel) that peaked in
intensity in 1998/99 (described in detail by K98 and K03)
when it rivalled in $V$ the brightness of the Mira but went unnoticeable in
$I_{\rm C}$.  This corresponds to the typical, non-TNR outbursts that
essentially all symbiotic stars have experienced several times in their
recorded photometric history; and (3) the violent, rapid and bright TNR
outburst of 2010. The latter overwhelmed the brightness of the Mira at
optical wavelengths, but only equalled it in the $K$ band (cf. data in
Table~4 with the long term $J$$H$$K$ light-curve of the Mira presented by
K98 and K03). Figure~2 presents a zoomed view on the earliest evolution
of the 2010 outburst in the $B$$V$$R_{\rm C}$$I_{\rm C}$$J$$H$$K$ bands.
The optical maximum was reached at $V$=7.1 on March 10.8 UT, and the
subsequent decline was very fast and characterized by $t^{V}_{2}$=5.9 and
$t^{V}_{3}$=24 days.  The decline was similarly fast  in T CrB,
V745 Sco, RS Oph and V3890 Sgr that showed $t_{2}$= 4, 5, 7 and 9 days,
respectively.  The $V$-band light-curve in Figure~1 shows a distinct knee at
day +47.  By analogy with RS Oph (cf.  Hachisu et al.  2006), it could mark
the end of the stable H-burning on the WD.

The outburst evolution seen in the $I_{\rm C}$ panel in Figure~1 could
appear in conflict with the expected underlying pulsation cycle of the Mira.
Indeed, the pulsation of the latter is known to be highly variable from
cycle to cycle (K98, Munari and Jurdana-Sepic 2002), with
puzzling sharp minima occurring at various pulsation phases (Kiziloglu and
Kiziloglu 2010; some of them are visible also in the light-curve of Figure~1
in 2007 and 2009), and that could be related to the unusual nature of the
Mira in V407 Cyg.  In fact, Miras of such a long pulsation period are usually
the central stars of OH/IR sources and their thick dust cocoon prevent them
from being visible in the optical.  As remarked by M90, the presence of the
hot and outbursting WD companion, could disturb the formation of the dust
cocoon and thus make the Mira in V407 Cyg visible at optical wavelengths.


\bsp

\clearpage

  \begin{table}
     \caption{Our optical photometry of V407 Cyg during the 2010 outburst.}
     \centering
     \includegraphics[width=8.0cm]{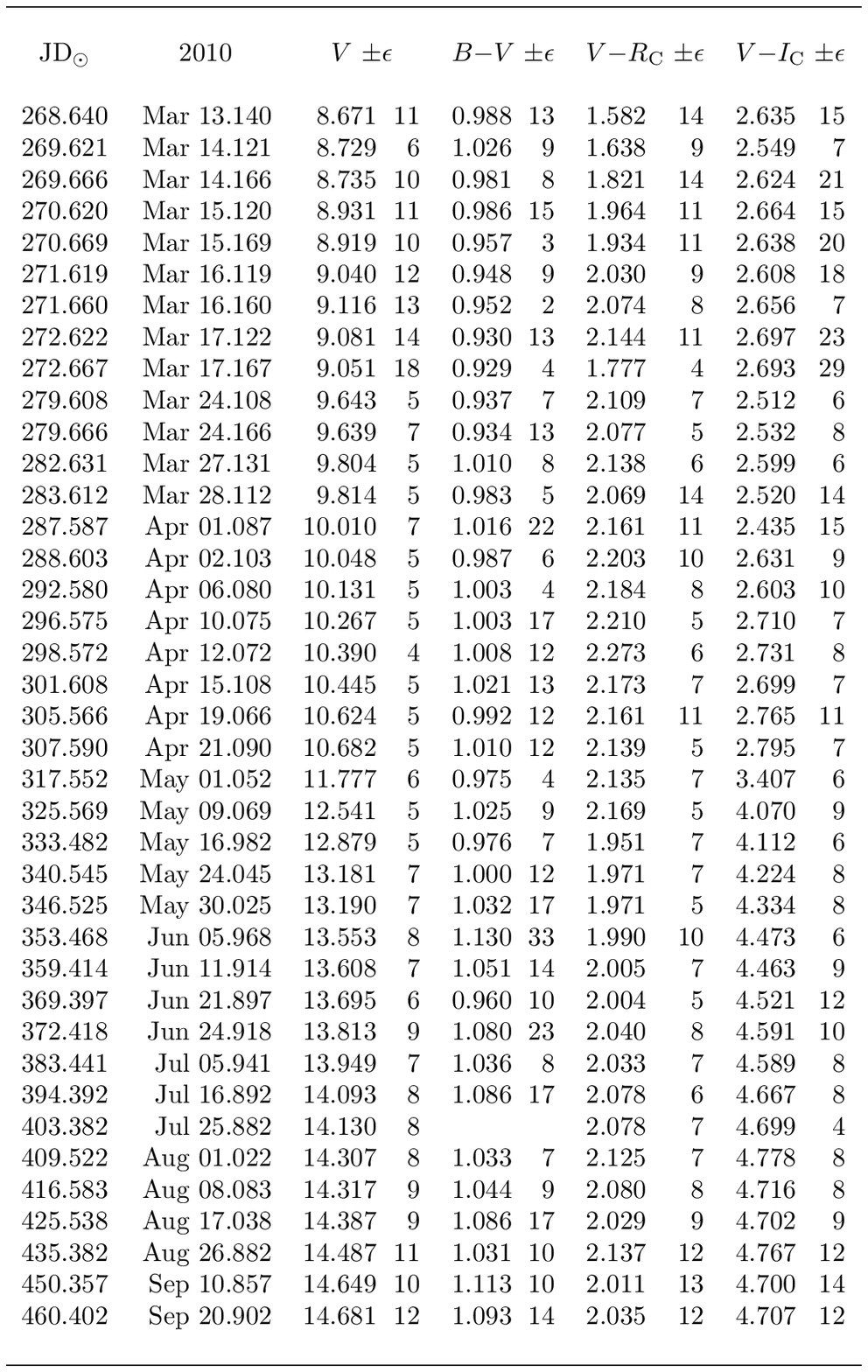}
     \label{tab1}
  \end{table}

  \begin{table}
     \caption{Journal of our optical spectroscopic observations of V407 Cyg
     during the 2010 outburst and a few earlier epochs.}
     \centering
     \includegraphics[width=8.0cm]{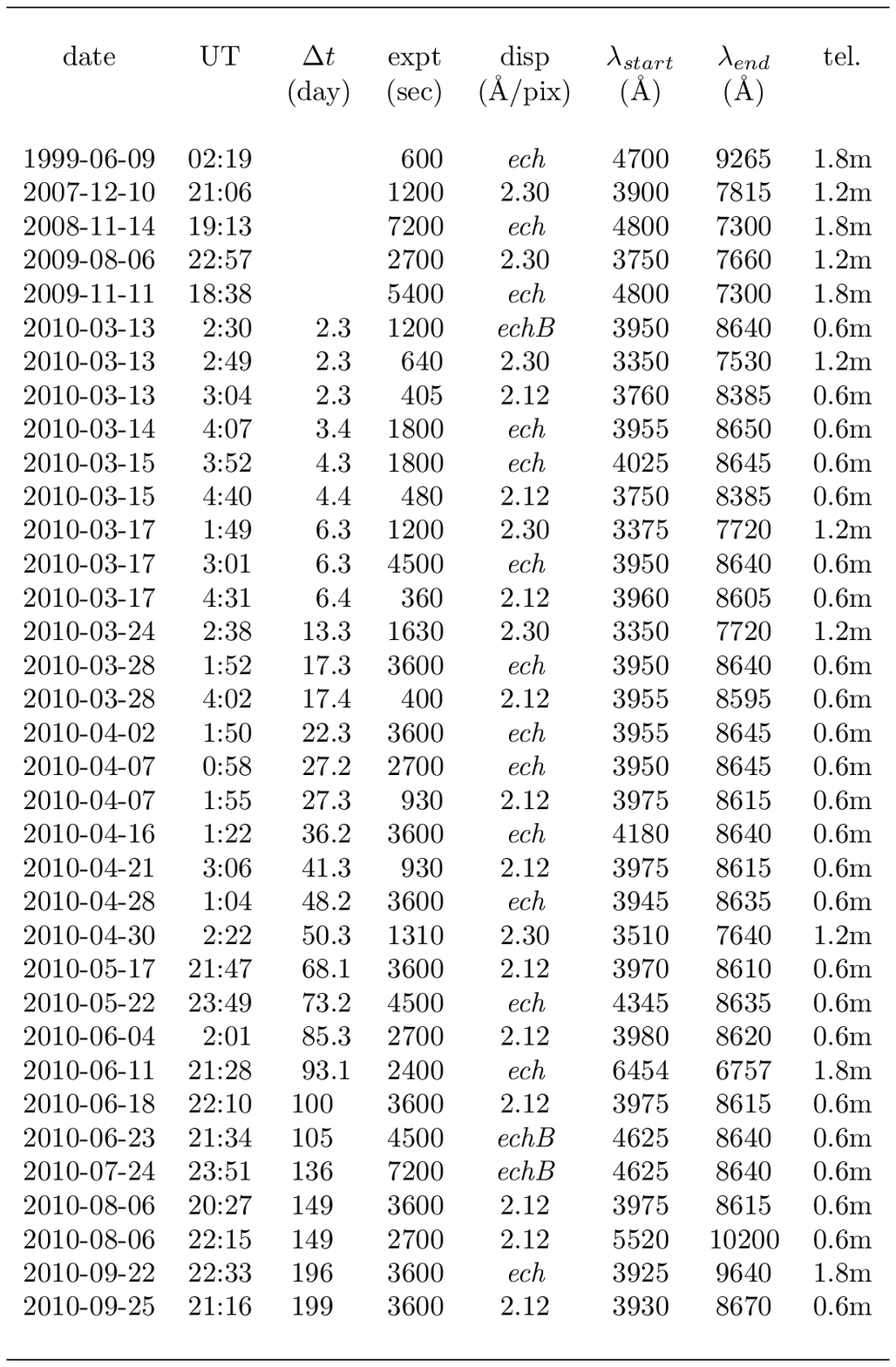}
     \label{tab2}
  \end{table}

  \begin{table}
     \caption{Journal of our infrared spectroscopic observations of V407 Cyg
     during the 2010 outburst. The integration times in brackets are for
     the J, H and K bands respectively}
     \centering
     \includegraphics[width=8.0cm]{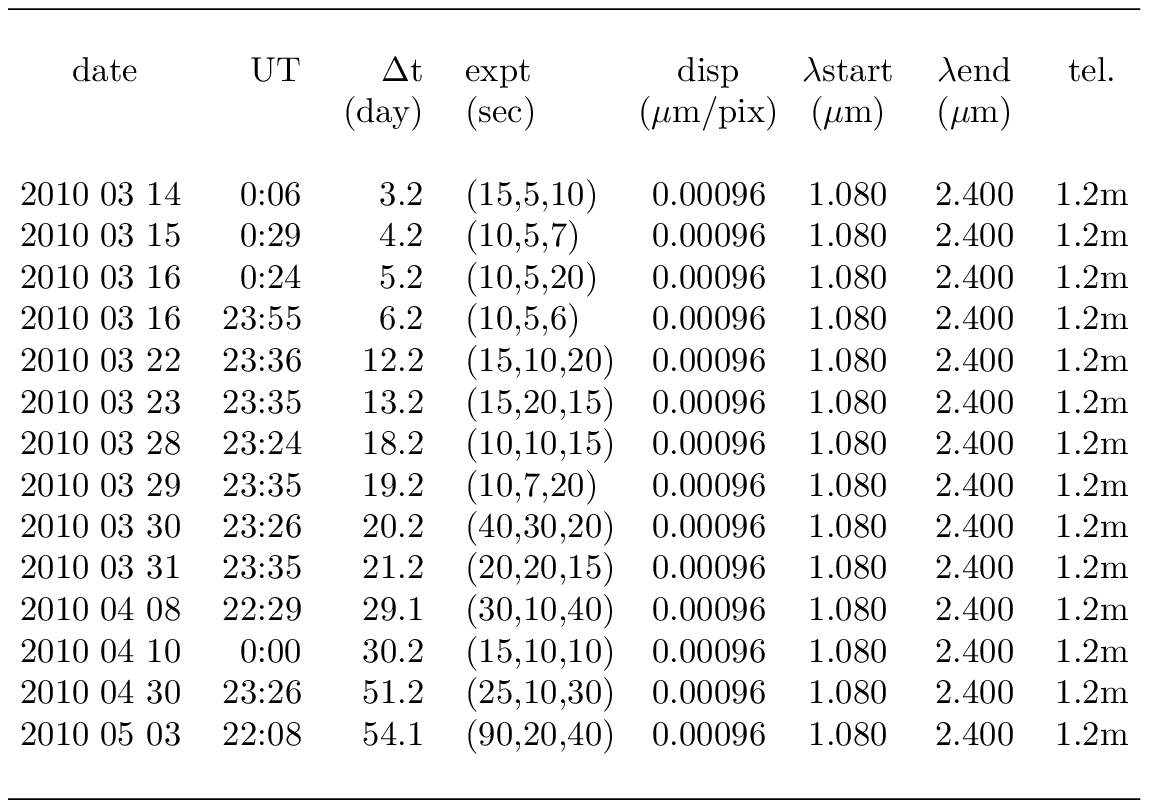}
     \label{tab2}
  \end{table}

  \begin{table}
     \caption{Our $J$$H$$K$ photometry of V407 Cyg
     during the 2010 outburst.}
     \centering
     \includegraphics[width=6.4cm]{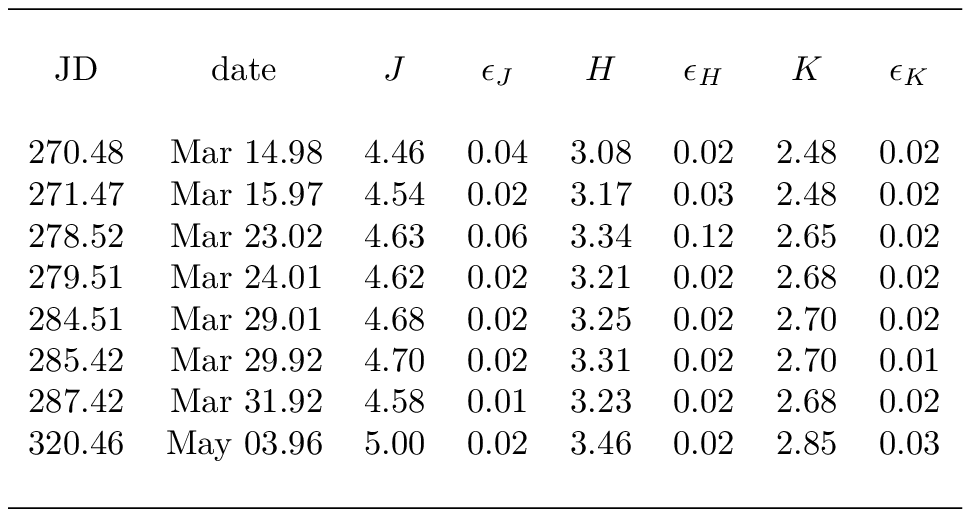}
     \label{tab1}
  \end{table}

\label{lastpage}

\end{document}